\documentclass[prb,superscriptaddress,twocolumn,floatfix]{revtex4-2}
\usepackage[utf8]{inputenc}
\usepackage[T1]{fontenc}
\newcommand{\ud}{\mathrm{d}}
\usepackage{amsmath}
\usepackage{amssymb}
\usepackage{newtxtext}
\usepackage[smallerops]{newtxmath}
\let\lambda\lambdaup
\allowdisplaybreaks
\usepackage{graphicx}
\graphicspath{{fig/}}
\usepackage{xcolor}
\usepackage[colorlinks=true,linkcolor=blue,citecolor=blue,urlcolor=blue]{hyperref}
\usepackage[capitalise]{cleveref}
\usepackage{multirow}
\usepackage[version=3]{mhchem}

\renewcommand{\Re}{\mathop{\mathrm{Re}}}
\renewcommand{\Im}{\mathop{\mathrm{Im}}}

\newcommand{\sgn}{\mathop{\mathrm{sgn}}}
\newcommand{\tr}{\mathop{\mathrm{tr}}}
\newcommand{\erf}{\mathop{\mathrm{erf}}}
\newcommand{\Jav}{\overline{J}}
\newcommand{\Qav}{\overline{Q}}
\newcommand{\lgi}{LiGa$_{0.95}$In$_{0.05}$Cr$_4$O$_8$}

\begin{document}

\title{Dynamics of the antiferromagnetic spin ice phase in pyrochlore spinels}

\author{Attila Szab\'{o}}
\address{ISIS Facility, Rutherford Appleton Laboratory, Harwell Campus, Didcot OX11 0QX, UK}
\address{Rudolf Peierls Centre for Theoretical Physics, University of Oxford, Oxford OX1 3PU, UK}
\author{Gøran J. Nilsen}
\address{ISIS Facility, Rutherford Appleton Laboratory, Harwell Campus, Didcot OX11 0QX, UK}
\address{Department of Mathematics and Physics, University of Stavanger, 4036 Stavanger, Norway}

\date{\today}

\begin{abstract}
Motivated by the classical spin-nematic state observed in the breathing pyrochlore spinel \lgi, we theoretically discuss spin dynamics in models of spin-lattice coupling in these materials.
Semiclassical dynamical simulations successfully recover the key features of inelastic neutron-scattering experiments on \lgi: a broad finite-energy peak alongside a continuum of scattering near the $(200)$ wave vector that extends from the elastic line to high energies. 
To interpret this result, we develop a small-fluctuation theory for the spin-ice-like nematic ground states, analogous to linear spin-wave theory for conventionally ordered magnets, which reproduces the numerical simulation results quantitatively.
In particular, we find that the inelastic peak is well explained by collective modes confined to ferromagnetic loops of the underlying nematic order.
In addition, we find a sharp, linearly dispersing mode in the dynamical structure factor, which originates in long-wavelength fluctuations of the nematic director: We believe identifying this mode will be an interesting target for future experiments on these materials. 
We also outline potential future applications of our methods to both pyrochlore spinels and other spin-nematic systems.
\end{abstract}

\maketitle

\section{Introduction}

In frustrated magnets, no single magnetic ground state is able to satisfy all competing interactions.
Such materials can show a wide range of exotic physical properties, including extensive ground-state degeneracy, fractional excitations, topological order, and other hallmarks of spin-liquid physics~\cite{Castelnovo2012SpinOrder,Balents2010SpinMagnets,Savary2017QuantumReview}.
Such behaviour is, however, usually limited to fine-tuned models:
In most real magnetic materials, additional (e.g., further-neighbour or spin--lattice) interactions and disorder tend to suppress spin liquids in favour of an ordered state or a spin glass~\cite{Clark2021QSLMaterials}.

Spinel pyrochlores of formula $AB_2$O$_4$, where the magnetic $B$ ions form a lattice of corner-sharing tetrahedra, exhibit a variety of such mechanisms. 
For example, the magneto-structural order of \ce{ZnCr2O4} at low temperature involves both a lowering of symmetry from cubic to tetragonal and a doubling of the unit cell, as well as a non-collinear spin arrangement~\cite{Ji2009ZnCr2O4}.
This complexity arises from the interplay between further-neighbour interactions and spin--lattice coupling~\cite{Tchernyshyov2002SpinPeierls,Conlon2010AbsentAntiferromagnet,Sushkov2005SpinPeierlsZnCr2O4}.
In the Zn$_{2-x}$Cd$_{x}$Cr$_2$O$_4$ solid solution, on the other hand, the introduction of bond disorder by chemical substitution produces an apparent spin glass ground state for even small values of $x$~\cite{Ratcliff2002FreezingZnCdSpinel}.

Perhaps the most intriguing mechanism for relieving frustration in spinels is due to spin--lattice coupling. 
The simplest approach to model this interaction is to consider a bond-length-dependent Heisenberg model coupled to phonons that modulate the bond lengths independently:
Integrating out the phonons then results in the bilinear-biquadratic (BLBQ) effective Hamiltonian~\cite{Kittel1960Biquadratic,Tchernyshyov2002SpinPeierls,Penc2004HalfMagnetization}
\begin{equation}
    \mathcal H = \sum_{\langle ij\rangle} \big[J \vec s_i\cdot\vec s_j - Q (\vec s_i\cdot\vec s_j)^2\big].
    \label{eq: Hamiltonian no breathing}
\end{equation}
In the large-$S$ limit (a justified approximation for the $S=3/2$ Cr$^{3+}$ ions), the pure Heisenberg model $Q=0$ is a classical spin liquid~\cite{Moessner1998PropertiesAntiferromagnet}.
Spin--lattice coupling introduces a finite $Q>0$, which causes the model to develop collinear nematic order at $T\simeq Q$~\cite{Shinaoka2014PhaseDiagram}.
This order, however, retains residual frustration: 
Once all spins align collinearly along a nematic director to optimise the $Q$ term, the $J$ term is equivalent to nearest-neighbour spin ice, which is optimised by exponentially many two-up-two-down states~\cite{Anderson1956OrderingFerrites}.
While the BLBQ model has been very successful in describing the magnetisation plateaux of several pyrochlore spinels~\cite{Penc2004HalfMagnetization,Miyata2011ZnCr2O4.600T,Miyata2012ZnCr2O4,Miyata2014MgCr2O4,Nakamura2014HgCr2O4},
longer-range interactions not captured by~\eqref{eq: Hamiltonian no breathing} cause most of them to exhibit full magnetic ordering rather than the predicted nematic spin-ice state.
In particular, the BLBQ model decouples the length modulation of different bonds around the same site:
in the more realistic \textit{site-phonon model}~\cite{Bergman2006PyrochloreDegeneracyBreaking,Aoyama2019SitePhonon}, phonons mediate further-neighbour multi-spin couplings that explain such orders as the plateau phases of \ce{CdCr2O4} and \ce{HgCr2O4}~\cite{Nakamura2014HgCr2O4}. 

In the past decade, much of the attention on pyrochlore spinels has shifted to the breathing pyrochlores $AA'$Cr$_4$O$_8$, where the ordering of the $A$ and $A'$ cations cause translation-inequivalent (up- and down-pointing) tetrahedra in the pyrochlore lattice to have different sizes and hence exchange couplings. 
High-temperature susceptibility measurements on the two best-known materials in the family, \ce{LiGaCr4O8} and \ce{LiInCr4O8}, indicate a ratio of Heisenberg couplings $J'/J$ of around 0.6 and 0.1 between the inequivalent tetrahedra, respectively~\cite{Okamoto2013BreathingSpinel}.
At low temperatures, both materials show magneto-structural ordering driven by spin--lattice coupling~\cite{Nilsen2015LiInMagnetostructural,Saha2016LiGaMagnetodielectric,Aoyama2019SitePhonon}, alongside phase separation due to site disorder.

Mixing Ga and In on the $A'$ site, however, quickly suppresses this ordering, with a possible gapped spin liquid on the In-rich side of the phase diagram, and a spin glass on the Ga-rich side~\cite{Okamoto2015PhaseDiagram}.
For \lgi\ (with $J'/J$ close to that of \ce{LiGaCr4O8}), however, neutron-diffraction and specific-heat measurements indicate that the nematic state predicted by the BLBQ model~\eqref{eq: Hamiltonian no breathing} is stabilised~\cite{Wawrzynczak2017LGI}.
This indicates that the bond disorder introduced by the Ga--In mixing is weaker than the leading biquadratic interaction generated by spin--lattice coupling, allowing nematic ordering, but it is strong enough to disrupt full ordering, stabilising instead a glassy spin-ice state~\cite{Shinaoka2014PhaseDiagram}, similar to dipolar spin ice~\cite{Melko2004MonteModel}.
Inelastic-neutron-scattering experiments on \lgi~\cite{Tanaka2018InelasticNeutron} found a broad peak in the dynamical structure factor $\mathcal{S}(q,\omega)$ at $\hbar\omega \approx 5.5$~meV and wave vector $q \approx 1.6$~\AA$^{-1}$ [corresponding approximately to the (200) reciprocal lattice vector for the cubic lattice constant $a_0=8.25$~\AA], tentatively ascribed to excitations localised on antiferromagnetic hexagon loops.

In this paper, we explore the dynamics of \lgi\ and the nematic state of breathing pyrochlores in general.
Our main focus will be the classical BLBQ model, generalised to the breathing lattice:
\begin{equation}
    \mathcal H = \sum_{\langle ij\rangle\in\uparrow} \big[J \vec s_i\cdot\vec s_j - Q (\vec s_i\cdot\vec s_j)^2\big] + \sum_{\langle ij\rangle\in\downarrow} \big[J' \vec s_i\cdot\vec s_j - Q' (\vec s_i\cdot\vec s_j)^2\big],
    \label{eq: Hamiltonian}
\end{equation}
where $\uparrow,\downarrow$ stand for up and down tetrahedra, respectively.
For brevity, we will use units in which the spin magnitude $|\vec s_i|$ is 1~%
\footnote[11]{If $S\neq1$, our dynamical results remain applicable if $J,Q$ are replaced with $JS,QS^3$. The powers of $S$ are determined by the number of spin operators in the effective Landau--Lifshitz field $\vec B_i = -\partial \mathcal H / \partial\vec s_i$.}
and introduce $\Jav=(J+J')/2, \Qav=(Q+Q')/2$. 
In \cref{sec: dynamics}, we discuss simulations of the semiclassical Landau--Lifshitz dynamics under~\eqref{eq: Hamiltonian}. The numerically obtained dynamical structure factor is dominated by a broad peak at $\hbar\omega=16\Qav$, a sharp linearly dispersing mode at small $q,\omega$, and a weak continuum extending up to about $4\Jav$. 
We discuss the fate of these features in the presence of Landau--Lifshitz damping and structural disorder;
in \cref{sec: lswt}, we explain them in terms of linear ``spin-wave'' theory around the disordered spin-ice ground states of~\eqref{eq: Hamiltonian}.
We find that the $16\Qav$ peak is caused by out-of-phase precession around \textit{ferromagnetic} loops of spins,
while the linearly dispersing feature originates in long-wave fluctuations of the nematic director.
We perform the same analysis for the more complex site-phonon model as well:
we find that all qualitative features of the BLBQ dynamics survive, albeit with strong effective disorder and an approximately halved effective $Q$.
We discuss our findings in the context of experimental results in \cref{sec: discussion}. 

\section{Dynamical simulations}
\label{sec: dynamics}

We drew initial configurations from the thermal ensemble $e^{-\beta \mathcal H}$ using single-spin-flip Metropolis Monte Carlo on $16\times16\times16$ pyrochlore cubic unit cells (65\,536 spins). Similar to Ising spin ice~\cite{Melko2004MonteModel}, we expect Monte Carlo dynamics to slow down substantially in the nematic phase. Therefore, to avoid getting stuck in local minima that do not satisfy the ice rules, we used simulated annealing from a temperature well above the nematic transition [at least $2\max(Q,Q')$] down to $T=0.01\overline{J}$, well below the transition in every case, where we performed all dynamical simulations.

\begin{figure}
    \centering
    \includegraphics{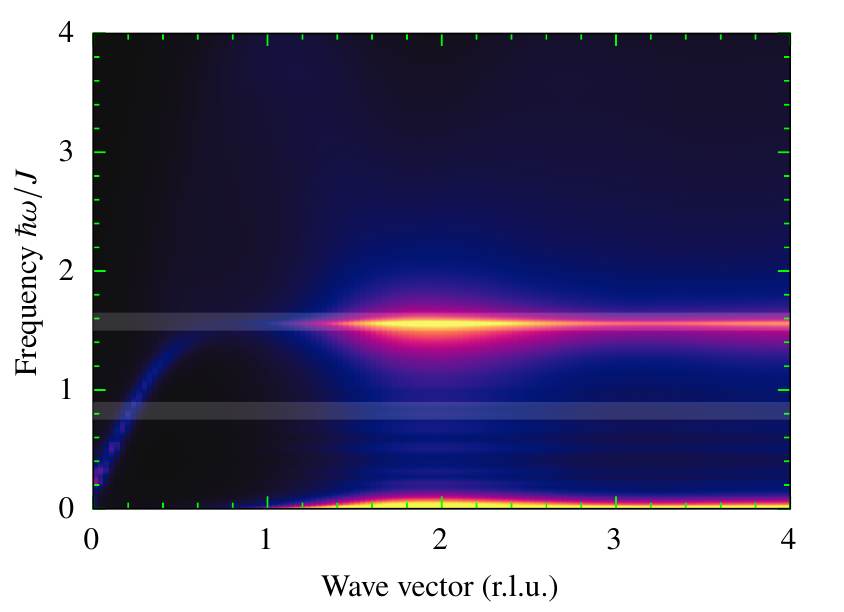}
    \caption{Simulated powder-averaged dynamic structure factor for $J=J', Q=Q'=0.1J, \alpha=0.01$. The pattern is dominated by a sharp maximum at $\hbar\omega\approx1.6J$ and a linearly dispersing mode that connects this maximum to the origin. }
    \label{fig: powder pattern}
\end{figure}

We then computed the time evolution of the initial spin configurations under the stochastic Landau--Lifshitz dynamical equation
\begin{align}
    \hbar\frac{\ud \vec s_i}{\ud t} &= \vec s_i \times (\vec B_i + \vec b_i) - \alpha \vec s_i \times\vec s_i \times (\vec B_i + \vec b_i),
    \label{eq: Landau Lifshitz}
\end{align}
where $\vec B_i = -\partial \mathcal{H} / \partial \vec s_i$ is the effective field acting on spin $\vec s_i$ and $\vec b_i$ is a stochastic field satisfying the fluctuation--dissipation relation
\begin{align}
    \langle b_i^\alpha(t)\rangle &= 0; &
    \langle b_i^\alpha(t) b_j^\beta(t')\rangle &= 2D \delta_{ij} \delta^{\alpha\beta} \delta(t-t'); \nonumber\\*
    &&D &= \frac{\alpha}{1+\alpha^2} k_\mathrm{B} T \hbar.
    \label{eq: fluctuation dissipation}
\end{align}
We note that the sign of the precession term in~\eqref{eq: Landau Lifshitz} is flipped compared to its usual presentation~\cite{Lakshmanan2011FascinatingWorld}, as we take the negative gyromagnetic ratio of the electron into account through the definition of $\vec B_i$. Details of the numerical method are described in Appendix~\ref{app: dynamics}; we benchmarked the simulations by ensuring that thermodynamic properties, such as the average energy, match the Monte Carlo results within statistical error.

We first focused on the case $J=J', Q=Q'=0.1J, \alpha=0.01$. The powder-averaged dynamical correlation function $S(q,\omega)$ is plotted in Fig.~\ref{fig: powder pattern}.
Similar to the experimental powder neutron-scattering pattern of Ref.~\cite{Tanaka2018InelasticNeutron}, we see a prominent scattering maximum at energy transfer $\hbar\omega\approx 1.6J$, with the highest intensity at $q\approx 4\pi/a_0$. 
In addition, we observe a linearly dispersing branch, extending from the origin to about the frequency of the intensity maximum.

\begin{figure*}[p]
    \centering
    \includegraphics{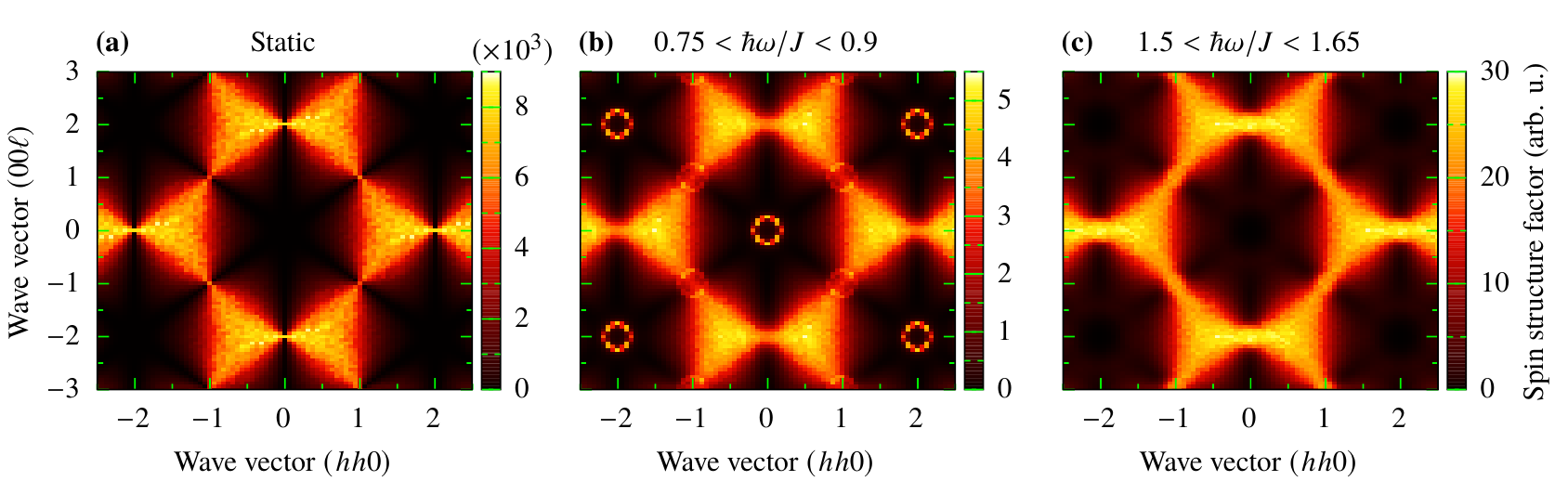}
    \caption{Static and frequency-window-integrated dynamical structure factors in the $(hh\ell)$ plane for the same parameters as in Fig.~\ref{fig: powder pattern}. The frequency windows used in (b,\,c) are highlighted with lighter background in Fig.~\ref{fig: powder pattern}. The ``arbitrary units'' are consistent across the three plots: the static structure factor, dominated by the $\omega=0$ component, is much larger than dynamical correlations, consistent with (nematic) ordering.}
    \label{fig: hhl plots}    
\end{figure*}

\begin{figure*}[p]
    \centering
    \includegraphics{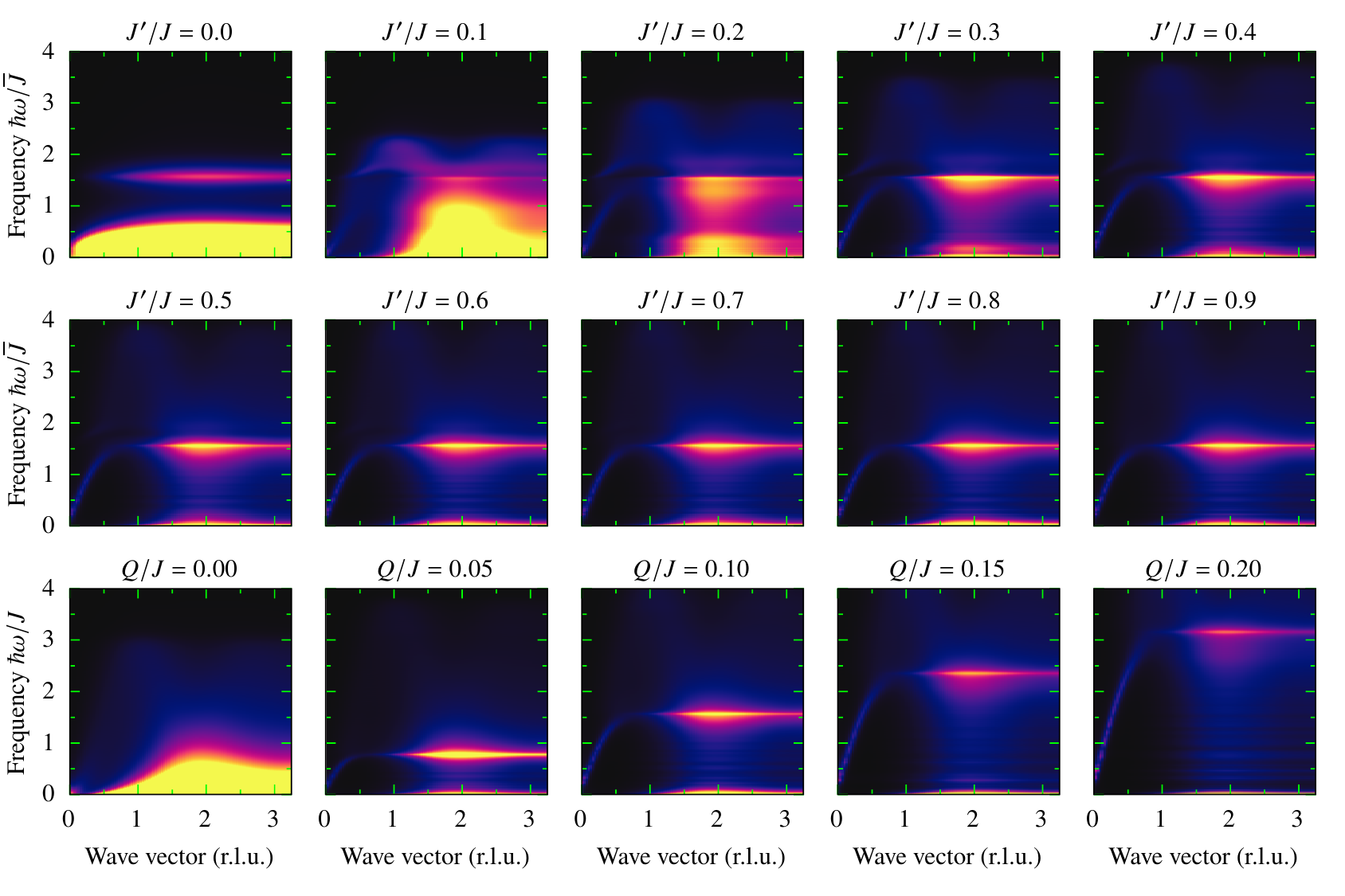}
    \caption{Simulated powder-averaged dynamic structure factor for several $J'/J$ at $Q/J=Q'/J'=0.1$ (top two rows) and for several $Q/J$ at $J=J', Q=Q'$ (bottom row). Except for $J'\ll J$ and $Q=0$, where the nematic ordering breaks down, the pattern is dominated by a broad feature at $16\overline Q$ and a linearly dispersing mode stretching from the origin to this feature. The colours indicate the same range of intensities in all panels except the last four; these are scaled as $I_\mathrm{max} \propto 1/Q$ to keep the total intensity of the $16Q$ feature visually constant.}
    \label{fig: all powder patterns}
\end{figure*}

We also plotted the static structure factor $S(q)$, as well as $S(q,\omega)$ integrated over two frequency windows, in Fig.~\ref{fig: hhl plots}. The static structure factor shows sharp pinch points in the pattern seen in Ref.~\cite{Henley2005Power-lawAntiferromagnets} for spin ice; this is expected as the effective Ising spins in the nematic order obey the same ice rules.
The same pinch points, albeit much broadened, are seen at finite $\omega$ as well; below the intensity maximum at $1.6J$, we also see sharp circular features corresponding to the linearly dispersing mode in Fig.~\ref{fig: powder pattern}.

As shown in Fig.~\ref{fig: all powder patterns}, these features of the scattering pattern are quite robust in parameter space.
While the model with $J'=Q'=0$ is qualitatively different from the symmetric case (it is made up of disconnected tetrahedra), most features of the latter are already recovered for $J/J'=0.2$ and the powder patterns at $J'/J\ge 0.4$ only differ in quantitative details.
Likewise, the scattering pattern of the pure Heisenberg model $Q=Q'=0$ is qualitatively different due to the lack of nematic ordering, but any finite $Q$ is enough to bring about both the linearly dispersing mode and a broad scattering maximum at $16\Qav$.

\begin{figure}
    \centering
    \includegraphics{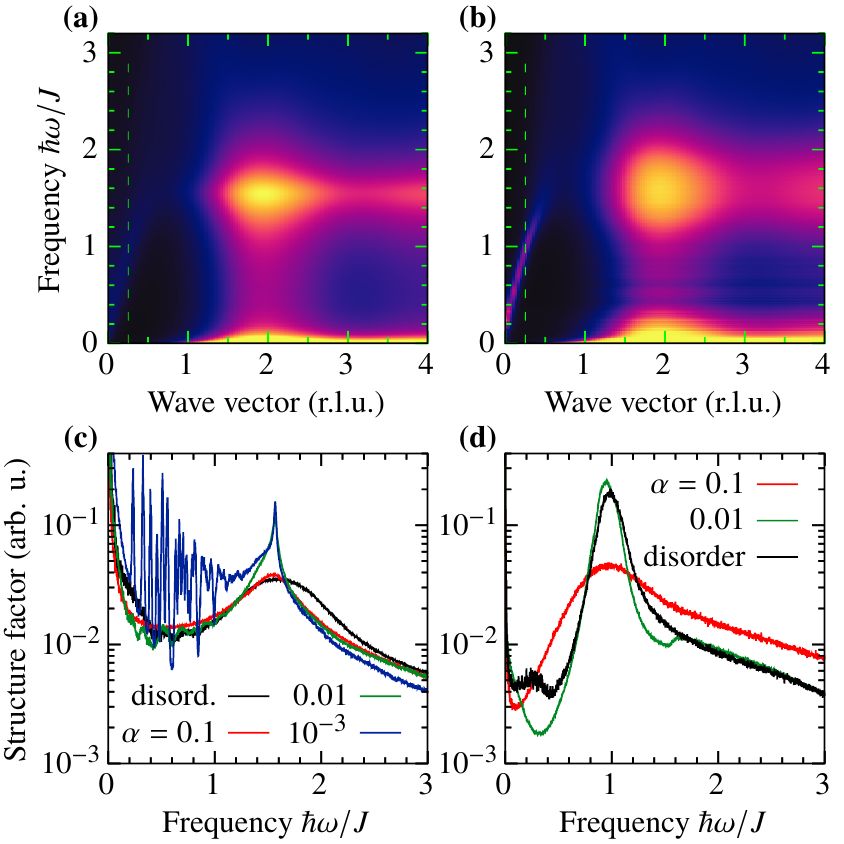}
    \caption{(a,\,b) Simulated powder-averaged dynamic structure factor for the same parameters as Fig.~\ref{fig: powder pattern} but with $\alpha=0.1$ (a) or 10\% Gaussian disorder in $J$ and $Q$ (b).
    (c) $q$-integrated structure factors for the disordered model and three values of $\alpha$.
    (d) Cuts of the powder pattern for the disordered model and two values of $\alpha$ at $q=\pi/(2a_0)$ [dashed green lines in panels (a,\,b)].
    The ``arbitrary units'' are not consistent between panels (c,\,d).
    }
    \label{fig: broadening}
\end{figure}

The intensity maximum at $16\Qav$ in our simulations is much sharper than the analogous experimental feature~\cite{Tanaka2018InelasticNeutron}.
This may either be because interactions with other dynamical degrees of freedom (e.g., magnon--magnon interactions) induce stronger damping than the Landau--Lifshitz damping term $\alpha=0.01$ used above, or because the exchange couplings $J,Q$ are disordered due to structural disorder or magnetoelastic distortions~\cite{Shinaoka2014PhaseDiagram}.
To distinguish these possibilities, we performed dynamical simulations at $J=J',Q=Q'=0.1J$ for two additional values of $\alpha=0.1,0.001$, as well as for a disordered model in which $J,Q$ for each bond is independently drawn from a Gaussian distribution of 10\% standard deviation relative to the mean.
The results are summarised in \cref{fig: broadening}. 
The $16\Qav$ feature is broadened by a similar amount both for $\alpha=0.1$ [\cref{fig: broadening}(a)] and on introducing disorder [\cref{fig: broadening}(b)]: the line shapes of the $q$-integrated structure factor [\cref{fig: broadening}(c)] appear Lorentzian and Gaussian, respectively, although the actual line shape is difficult to distinguish due to the background intensity.
The linearly dispersing mode, however, behaves qualitatively differently: it does not blur noticeably on introducing disorder but becomes broad and faint beyond the point of clear detection on increasing $\alpha$ [see also \cref{fig: broadening}(d)]. 
Lowering $\alpha$ causes the low-frequency structure factor to decompose into discrete normal modes of the finite simulation box, resulting in an array of sharp peaks in \cref{fig: broadening}(c).

\begin{figure}
    \centering
    \includegraphics{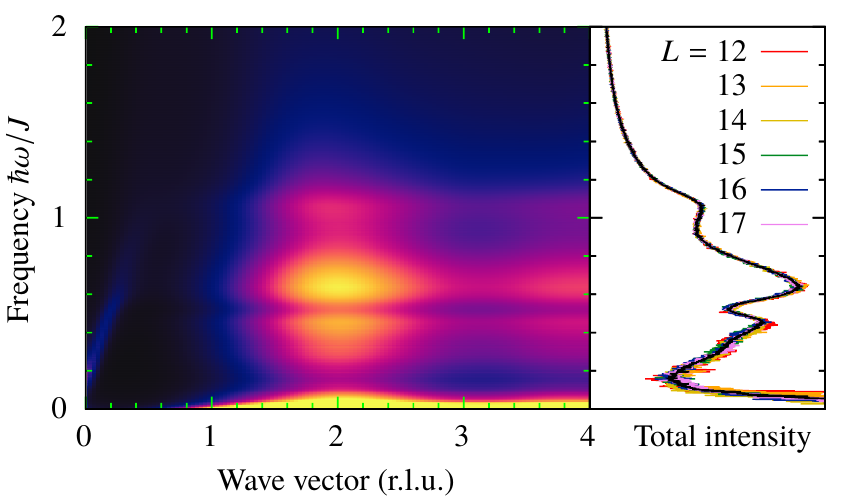}
    \caption{Simulated powder-averaged (left) and $q$-integrated (right) dynamic structure factor for the same parameters as Fig.~\ref{fig: powder pattern} in the site-phonon model~\eqref{eq: site-phonon Hamiltonian}. The powder pattern is averaged from simulations of clusters between $12^3$ and $17^3$ cubic unit cells; the integrated structure factor is shown for each cluster as well as the average. }
    \label{fig: site phonon}
\end{figure}

Finally, we considered dynamics under the site-phonon Hamiltonian
\begin{align}
    \mathcal H_\mathrm{sp} &= \sum_{\langle ij\rangle\in\uparrow} J \vec s_i\cdot\vec s_j + \sum_{\langle ij\rangle\in\downarrow} J' \vec s_i\cdot\vec s_j \nonumber\\*
    &-\frac12 \sum_i \sum_{j,k\sim i} \sqrt{Q_{ij}Q_{ik}} (\hat e_{ij} \cdot \hat e_{ik}) (\vec s_i \cdot\vec s_j)(\vec s_i \cdot\vec s_k),
    \label{eq: site-phonon Hamiltonian}
\end{align}
where the inner sum runs over all pairs $j,k$ of nearest neighbours of site $i$ ($j,k$ and $k,j$ are both counted),  $Q_{ij}=Q$ ($Q'$) if the bond $ij$ is part of an up (down) tetrahedron, and $\hat e_{ij}$ is the unit vector pointing from site $i$ to $j$.
Since this Hamiltonian has a fully ordered ground state~\cite{Bergman2006PyrochloreDegeneracyBreaking,Aoyama2019SitePhonon}, we emulated the glassy nematic order of \lgi\ by first preparing low-temperature states of the BLBQ Hamiltonian and annealing them under~\eqref{eq: site-phonon Hamiltonian} before measuring dynamical correlation functions. 
At $J=J', Q=Q'=0.1J$, we obtained the powder-averaged structure factors shown in Fig.~\ref{fig: site phonon}; see Appendix~\ref{app: site phonon gallery} for a wider range of parameters.
The general structure of the powder pattern remains unchanged and, in particular, the nematic state appears to be metastable even without quenched disorder.
However, the finite-frequency peak becomes much broader, and the peak frequency is reduced substantially, from $16Q$ to about $7Q$.
The linearly dispersing mode remains sharp, but its velocity is reduced, too.

Remarkably, we see a strong modulation of the intensity with frequency that appears to split the peak into a number of fringes.
Similar, albeit weaker, fringes have already appeared in the BLBQ model [cf.~Fig.~\ref{fig: broadening}(b)]: 
those are caused by the finite gap between normal modes in the linearly dispersing mode in a finite sample.
To rule out this origin for the modulation seen in the site-phonon model, we performed dynamical simulations on clusters of $L^3$ cubic unit cells for every $12\le L\le 17$.
After averaging the powder pattern for the different clusters, several fringes at low frequencies (where finite-size effects are the most pronounced) indeed disappear; however, the peak remains split into three.
Nevertheless, we expect that these fringes would be washed out either by stronger damping or quenched disorder, similar to the peak broadening seen in the BLBQ model in Fig.~\ref{fig: broadening}.

\section{Linear spin-wave theory (LSWT)}
\label{sec: lswt}

Spin dynamics in the nematically ordered phase consists of small thermal fluctuations around the equilibrium configuration, a spin-ice configuration with an arbitrary Ising axis. Without loss of generality, we choose this axis to be $\pm s^z$, so we can write 
\begin{equation}
    \vec s_i \simeq \left(\Re s_i^+, \Im s_i^+, S_i\sqrt{1-s_i^+s_i^-}\right),
    \label{eq: LSWT expansion}
\end{equation}
where the $S_i = \pm 1$ satisfy the ice rules, and $s_i^+ \sim \sqrt{T/\Jav} \ll 1$. Substituting this into~\eqref{eq: Hamiltonian} and expanding to quadratic order in $s^\pm$ gives
\begin{align}
    \mathcal H = \mathrm{const.} + \frac12 \sum_{ij} s_i^- H_{ij} s_j^+,
    \label{eq: Hamiltonian quadratic}
\end{align}
where the nonzero matrix elements $H_{ij}$ are
\begin{equation}
    H_{ij} =\begin{cases}
        J+J'+6(Q+Q') & i=j \\
        J - 2QS_iS_j & \langle ij\rangle \in \uparrow\\
        J' - 2Q'S_iS_j & \langle ij\rangle \in \downarrow
    \end{cases}
    \label{eq: Hamiltonian terms}
\end{equation}
in an ice-like arrangement of $S_i$.
Likewise, substituting~\eqref{eq: LSWT expansion} into the dynamical equation~\eqref{eq: Landau Lifshitz} (without the stochastic fields $\vec b_i$) and expanding to linear order gives (see Appendix~\ref{app: LSWT derivation})
\begin{equation}
    \hbar\frac{\ud s_i^+}{\ud t} = -(iS_i +\alpha) \sum_j H_{ij} s_j^+.
    \label{eq: Landau Lifshitz linearised}
\end{equation}
The dynamical modes of~\eqref{eq: Landau Lifshitz linearised} and their frequencies are given by the eigenvalue equation
\begin{equation}
    (S-i\alpha)H |r_a\rangle = \hbar\omega_a |r_a\rangle,
    \label{eq: lswt eigenvalue eq}
\end{equation}
where we introduce bra-ket notation for the vectors comprised of all $s_i^+$ and define for convenience the diagonal matrix $S$ with the Ising configuration $S_i$ along the diagonal.
For $\alpha=0$, all eigenfrequencies of~\eqref{eq: lswt eigenvalue eq} are real (see \cref{app: LSWT matrices}), as expected for energy-conserving dynamics near a ground state. Likewise, for $\alpha>0$, all modes decay exponentially.

\begin{figure}
    \centering
    \includegraphics{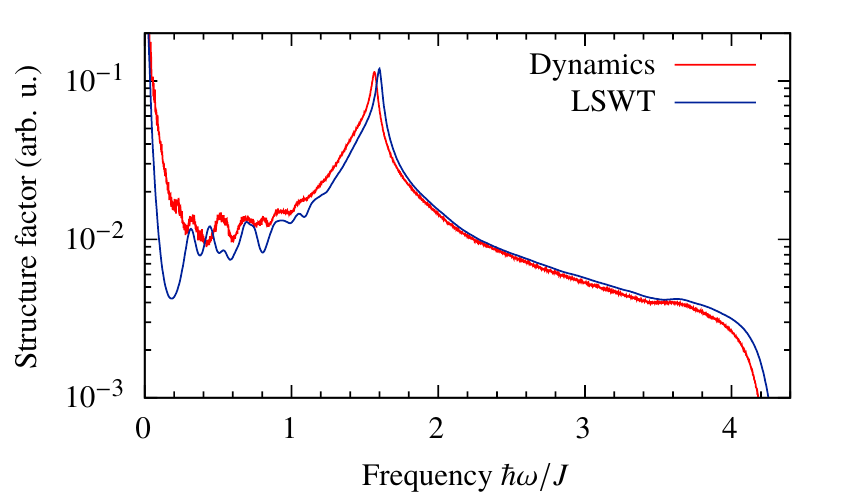}
    \caption{$q$-integrated structure factor obtained from dynamical simulations (red) and linear spin-wave theory (blue) for $J=J', Q=Q'=0.1J, \alpha=0.01$.}
    \label{fig: lswt intensity}
\end{figure}

We diagonalised~\eqref{eq: lswt eigenvalue eq} for a cluster of $12\times12\times12$ cubic unit cells (27\,648 spins), for both $\alpha=0$ and 0.01.
As explained in Appendix~\ref{app: LSWT dynamical structure factor}, these eigenvalues and eigenvectors can be used to compute the dynamical structure factor $S(q,\omega)$ within the linear-spin-wave approximation.
We find excellent quantitative agreement in the $q$-integrated structure factor (Fig.~\ref{fig: lswt intensity}) as well as the  powder pattern (not shown).
The two curves in Fig.~\ref{fig: lswt intensity} differ in two ways.
First, the low-frequency oscillations show a different pattern due to the different system sizes (and thus different low-frequency modes). 
Second, higher-frequency features of the LSWT spectrum are consistently shifted to slightly higher frequencies. This is due to spin-wave interactions, most of which can be accounted for in a simple mean-field picture: As the length $|\vec s_i|$ of spins is fixed to 1, transverse fluctuations cause $\langle s^z_i\rangle=:s_0$ to shorten, which renormalises the coefficients of~\eqref{eq: Landau Lifshitz linearised} as $J \mapsto Js_0, Q \mapsto Qs_0^3$~\cite{Tanaka2018InelasticNeutron,Note11}. From~\eqref{eq: static trace}, we estimate $s_0 = \sqrt{1-\langle s_i^+ s_i^-\rangle} \approx 0.9916$; scaling LSWT frequencies by a factor of $s_0^3$ indeed causes the $16Q$ peaks of the two curves to overlap perfectly.

In summary, spin waves give a full, quantitative account of the inelastic spin dynamics, and nonlinear effects affect the spin-wave spectrum very weakly at low temperatures.
In the following sections therefore, we will explain the salient features of the dynamical structure factor in terms of particular eigenmodes of the dynamical equation~\eqref{eq: Landau Lifshitz linearised}.

\subsection{Exact eigenmodes at $16\Qav$}
\label{sec: 16Q modes}

In the simulations, we see an accumulation of eigenfrequencies near $16\Qav$. More interestingly, a number (140 for the pattern of $S_i$ we used) of them is equal to $\pm 16\Qav$ \textit{within numerical accuracy} for $\alpha=0$.  We also observe that these exact eigenmodes live exclusively on up spins (for $\omega>0$) or down spins (for $\omega<0$).

Since the LSWT matrix $H$ only has on-site and nearest-neighbour matrix elements, we can decompose it into terms acting on a single tetrahedron only. These terms have the form
\begin{equation}
    H_\uparrow = \left(\begin{array}{cccc}
        J+6Q & J-2Q & J+2Q & J+2Q \\
        J-2Q & J+6Q & J+2Q & J+2Q \\
        J+2Q & J+2Q & J+6Q & J-2Q \\
        J+2Q & J+2Q & J-2Q & J+6Q 
    \end{array}\right),
    \label{eq: Hamiltonian on one tetrahedron}
\end{equation}
for up tetrahedra; for down tetrahedra, $J\to J', Q\to Q'$.
The first two and last two rows and columns of the matrix correspond to up and down spins, respectively. 
$H_\uparrow$ has two eigenvectors with eigenvalue $8Q$: they are orthogonal to both $(1,1,1,1)$ (i.e., they respect the ice rules) and the spin configuration of the tetrahedron. 
By enforcing these constraints on every tetrahedron, we can construct a number of eigenmodes of $H$; the corresponding eigenvalue is $8(Q+Q')=16\Qav$ as every spin belongs to one up and one down tetrahedron.

To obtain an eigenvector of the dynamical matrix $SH$ from this construction, we need them to be eigenvectors of $S$ as well, so they must be constrained to up ($S=+1$) or down ($S=-1$) spins in the nematic Ising configuration. On each tetrahedron, there are two configurations that obey all of these requirements: out-of-phase fluctuations of either the two up or the two down spins. We can build joint eigenstates of $H$ and $S$ from these by following closed loops of up or down spins and giving nearest neighbours out-of-phase fluctuations. In periodic boundary conditions, the loops always close, so the resulting fluctuation vectors are exact eigenvectors of both $H$ and the dynamical matrix $SH$. 
Therefore, they are also eigenmodes of the damped dynamics~\eqref{eq: lswt eigenvalue eq} with complex frequency $\hbar\omega = 16(\pm1-i\alpha)\Qav$,
as we also found in exact diagonalisation.

\begin{figure}
    \centering
    \includegraphics{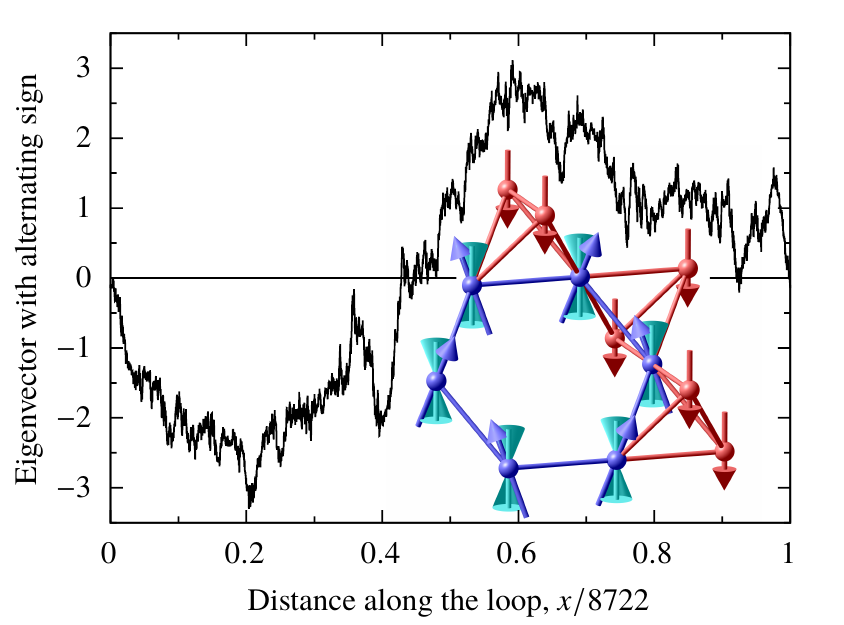}
    \caption{An eigenvector $|r\rangle$ of the $\alpha=0$ dynamical matrix with frequency $\hbar\omega/J=-1.6+3.01\times10^{-8}$, restricted on the longest closed loop of down spins, where 93\% of the statistical weight falls. 
    Inset: illustration of the exact $16\Qav$ eigenmode on the shortest possible ferromagnetic loop (blue atoms). Magnetic moments around the loop (blue arrows) precess around their equilibrium Ising direction (cones); the fluctuations of nearest neighbours are out of phase.
    }
    \label{fig: approx eigenstate}
\end{figure}

Since each spin has precisely two neighbours of the same spin in a spin-ice configuration (one on each tetrahedron), the loops that give rise to these states are uniquely defined. By constructing them on simulated ice configurations, we found that their lengths have a very broad distribution: a few loops cover almost all spins, while the remaining ones form very small loops, often as small as a single hexagon. The exact eigenmodes on the latter resemble the ``weathervane modes'' proposed in Ref.~\cite{Tanaka2018InelasticNeutron} (inset of Fig.~\ref{fig: approx eigenstate}).

The exact eigenmodes described above, however, do not account for the full intensity of the $16\Qav$ peak in the dynamical structure factor, or the high density of LSWT eigenmodes \textit{near} this frequency.
On a long loop, however, we can consider ``excited loop states,'' in which the exact eigenmode with alternating phases is modulated with a standing wave along the loop. Locally, this pattern is very similar to the exact eigenmode, thus we expect them to be eigenmodes to a good approximation, with a frequency very close to $\pm16\Qav$. 
A few numerically obtained eigenmodes of $SH$ follow this pattern closely (Fig.~\ref{fig: approx eigenstate}) and most of those near $\omega=\pm16\Qav$ show similar features, albeit obscured somewhat by local interference between different loops.
Furthermore, the fact that these modes live on loops explains the singular cusp, characteristic of \textit{one-dimensional} van Hove singularities, in the structure factor.

\subsection{Low-frequency dispersive modes}

\begin{figure}
    \centering
    \includegraphics{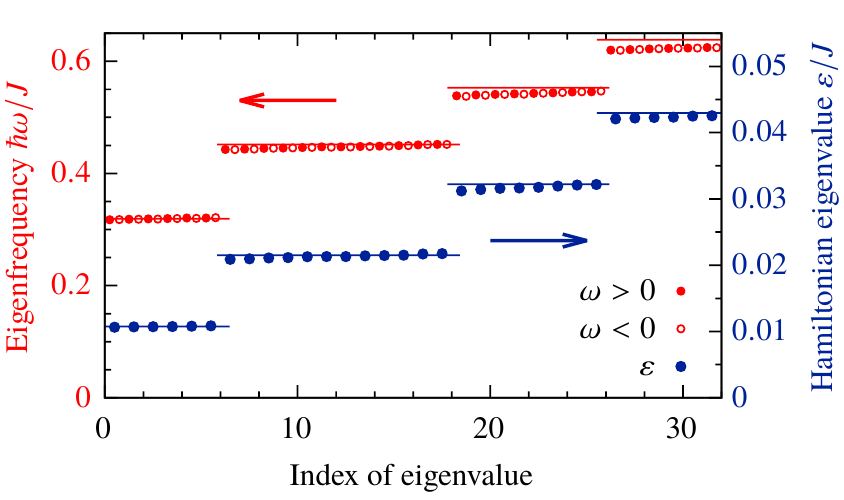}
    \caption{Lowest-magnitude eigenvalues of $H$ (blue) and $SH$ (red). The spectrum forms approximate multiplets of multiplicity $6,12,8,6,\dots$ (one sequence each for $\pm\omega$). The eigenvalues in the $n$th multiplet scale as $\omega\propto\pm\sqrt{n}$ and $\varepsilon\propto n$ (indicated by the horizontal lines) to a good approximation.}
    \label{fig: low freq modes}
\end{figure}

For $\alpha=0$, we found that the lowest-magnitude eigenvalues of both $H$ and the dynamical matrix $SH$ organise themselves in approximate multiplets (Fig.~\ref{fig: low freq modes}).
Their multiplicities match those of the reciprocal lattice vectors $\{100\},\{110\},\{111\},\{200\},\dots$ of the cubic simulation box, while the eigenvalues of $SH$ and $H$ scale with the same wave vectors as $\omega\propto\pm |k|$, $\varepsilon\propto k^2$, respectively. 
This indicates a linearly dispersing dynamical mode, consistent with the dynamical simulations.
The corresponding eigenvectors do not show a particularly high overlap with the plane waves $|k\rangle$, but rather with the vectors
\begin{equation}
    S|k\rangle = \sum_i S_i e^{i\vec k\cdot \vec r_i} |\text{site $i$}\rangle.
\end{equation}
At $k=0$, this mode corresponds to rotating the Ising axis of the nematic order; for small $k$, it captures long-wavelength fluctuations of the director, which we anticipate to cost little energy.

To explain these findings, we first note that both $|k\rangle$ and $S|k\rangle$ are approximate eigenmodes of $H$ for small $k$. The first closely resembles the all-in-all-out configuration $(1,1,1,1)$ on each tetrahedron, which is an eigenvector of the single-tetrahedron Hamiltonian~\eqref{eq: Hamiltonian on one tetrahedron} with eigenvalue $4J+8Q$. Since each spin belongs to one up and one down tetrahedron, these contributions add up to give
\begin{equation}
    H|k\rangle = \underbrace{[4(J+J')+8(Q+Q')]}_E |k\rangle + O(k^2).
\end{equation}

In the limit $J\to 0$, $S|k\rangle$ is an \textit{exact} eigenvector of $H$: the matrix elements of $SHS$ have no factors of $S_i$, so it is translation symmetric and its eigenvectors are plane waves~%
\footnote{Strictly speaking, we get four bands corresponding to the four fcc sublattices of the pyrochlore lattice. However, the eigenvector of the lowest band is $(1,1,1,1) + O(k^2)$, so we can treat the exact low-energy eigenvectors as plane waves to leading order. The higher-order corrections are responsible for the dispersive features at other $\Gamma$ points seen in Fig.~\ref{fig: hhl plots}(b).}
with eigenvalue $\varepsilon(k) = \Qav k^2/4 + O(k^4)$. 
A finite $J$ adds disordered terms to $SHS$ that penalise fluctuations proportional to the spin configuration $S_i$ on each tetrahedron. However, the weight of such fluctuations is only $O(k^2)$ as the plane wave $|k\rangle$ is proportional to $(1,1,1,1)+O(k)$ on each tetrahedron.
That is, even if $J\gg Q$ causes these fluctuations to be completely projected out, the resulting eigenmode of $SHS$ is still $|k\rangle$ up to $O(k^2)$ corrections. That is,
\begin{align}
    SHS|k\rangle &= \varepsilon(k)[|k\rangle + O(k^2)], \nonumber\\*
    H(S|k\rangle) &= \varepsilon(k)[S|k\rangle + O(k^2)]
\end{align}
where $\varepsilon(k)\propto k^2$.

Now, up to $O(k^2)$ corrections, the dynamical equation~\eqref{eq: lswt eigenvalue eq} can be written for a mode $|r\rangle = a S|k\rangle + b|k\rangle$ as
\begin{equation}
    \hbar\omega \binom{a}{b} = \left(\begin{array}{cc}
        -i\alpha\varepsilon(k) & E \\
        \varepsilon(k) & -i\alpha E
    \end{array}\right) \binom{a}{b}.
    \label{eq: low freq secular eq}
\end{equation}
For $\alpha=0$, we obtain the eigenmodes
\begin{subequations}
\begin{align}
    \hbar\omega&=\pm \sqrt{\varepsilon(k)E} 
    \label{eq: low frequency omega}\\*
    |r\rangle &= S|k\rangle \pm \sqrt{\varepsilon(k)/E}\, |k\rangle.
    \label{eq: low frequency mode}
\end{align}
\end{subequations}
The numerically obtained $\varepsilon$ and $\omega$ plotted in Fig.~\ref{fig: low freq modes} match~\eqref{eq: low frequency omega} closely. Since $\varepsilon(k)\ll E$, the modes~\eqref{eq: low frequency mode} are dominated by $S|k\rangle$, but the small admixture of $|k\rangle$ is enough to yield a visible dispersing mode, as it is not diffuse in $k$-space.

For $\alpha\neq 0$, the complex eigenfrequencies of~\eqref{eq: low freq secular eq} are
\begin{equation}
    \hbar\omega = -i\alpha\frac{\varepsilon(k)+E}2 \pm \sqrt{\varepsilon(k)E - \alpha^2\left(\frac{\varepsilon(k)-E}2\right)^2}.
    \label{eq: low freq damped}
\end{equation}
Even for small $\alpha$, the decay rate $\Im\omega$ is independent of $k$, thus coherent oscillations at the longest wavelengths are always disrupted (the square root in~\eqref{eq: low freq damped} becomes imaginary, indicating purely decaying modes). 
For $\alpha=0.1$, this becomes the case for most values of $k$, that is, the linearly dispersing modes blur completely, as seen in the dynamical simulations.
By contrast, we expect that they remain robust against disorder: 
since they are dominated by long-wave modulations of the nematic director, they are only sensitive to coarse-grained averages of the exchange couplings $J,Q$, which are affected far less by disorder.

\subsection{Site-phonon model}

Finally, we consider the site-phonon Hamiltonian~\eqref{eq: site-phonon Hamiltonian}.
Expanding the Hamiltonian to quadratic order is somewhat more complicated than in the BLBQ case, and is described in detail in \cref{app: site phonon quadratic}.
We find that, despite the further-range quartic terms in~\eqref{eq: site-phonon Hamiltonian}, the matrix $H$ in~\eqref{eq: Hamiltonian quadratic} only contains nearest-neighbour terms.
In a spin-ice configuration, the coefficient of $s_i^-s_j^+$ becomes
\begin{equation}
    H_{ij} = J + Q-\sqrt{QQ'} - QS_iS_j + \sqrt{QQ'} \frac{S_iS_{j'}+S_jS_{i'}}2
    \label{eq: site phonon H matrix}
\end{equation}
if the bond $ij$ is on an up tetrahedron, where $j',i,j,i'$ are consecutive sites along a $\langle 110\rangle$ chain (cf.~Fig.~\ref{fig: site labels}); for a down tetrahedron, $J\to J', Q\leftrightarrow Q'$;
the diagonal terms $H_{ii}$ are determined by the constraint $H|S\rangle =0$ imposed by spin-rotation symmetry.
The first three terms of~\eqref{eq: site phonon H matrix} only renormalise $J,J'$:
unless $J'\ll J$ or $Q$ is large compared to $J$, this does not affect the stability of the dynamics.
The fourth term in analogous to the $Q$ term of~\eqref{eq: Hamiltonian terms}, but is halved in magnitude:
this accounts for most of the reduction in the inelastic-peak frequency.

The last term depends on spins outside of the bond, so it acts as disorder on top of this renormalised BLBQ quadratic Hamiltonian.
For $Q\approx Q'$, its magnitude is comparable to the renormalised $Q$ term, so it is expected to strongly broaden the finite-frequency peak, as indeed seen in Fig.~\ref{fig: site phonon}. 
To account for the remaining discrepancy in the peak frequency, we consider a simple mean-field picture, where we replace the last two terms with
\begin{equation*}
    -S_iS_j \left(Q - \sqrt{QQ'} \frac{\langle S_iS_{i'} \rangle + \langle S_jS_{j'} \rangle}{2}\right).
\end{equation*}
In nearest-neighbour spin ice at zero temperature, the correlator of two spins in this position is $\langle S_iS_{i'} \rangle = \langle S_jS_{j'} \rangle \approx 0.0883$;
at $Q=Q'$, this predicts a further renormalisation of $Q$ that brings the peak down from $8Q$ to $\approx7.3 Q$, in good quantitative agreement with Fig.~\ref{fig: site phonon}.

A detailed account of the splitting of the renormalised $16\Qav$ peak is beyond the scope of this work. 
However, we speculate that due to the strong but discrete [the last term of~\eqref{eq: site phonon H matrix} can only be $\pm \sqrt{QQ'}$ or 0] disorder, the eigenmodes living on long ferromagnetic chains discussed in Sec.~\ref{sec: 16Q modes} break up into short segments with equal disorder terms, leading to three peaks.

\section{Discussion}
\label{sec: discussion}

To summarise, dynamical simulations of the bilinear-biquadratic Hamiltonian~\eqref{eq: Hamiltonian} on the breathing pyrochlore lattice with $J'/J\gtrsim 0.3$ show a dynamical structure factor made up of three components:
(1) a broad inelastic peak at $16\Qav$,
(2) a sharp linearly dispersing mode, and
(3) a broad, weakly dispersing continuum extending to about $4\Jav$.
We accounted for this spectrum quantitatively in terms of small fluctuations around the spin-ice-like ground states of the nematically ordered model, similar to linear spin-wave theory around conventionally ordered magnets.
In particular, feature (1) is due to collective spin precession around long \textit{ferromagnetic} loops, while feature (2) originates in long-wavelength fluctuations of the nematic director.
We developed a similar small-fluctuation theory for the more accurate site-phonon model~\cite{Bergman2006PyrochloreDegeneracyBreaking,Aoyama2019SitePhonon}, which shows the same qualitative features, albeit with a renormalised dispersion relation: in particular, the position of the finite-frequency peak is renormalised down to about $7\Qav$.

\begin{figure}
    \centering
    \includegraphics{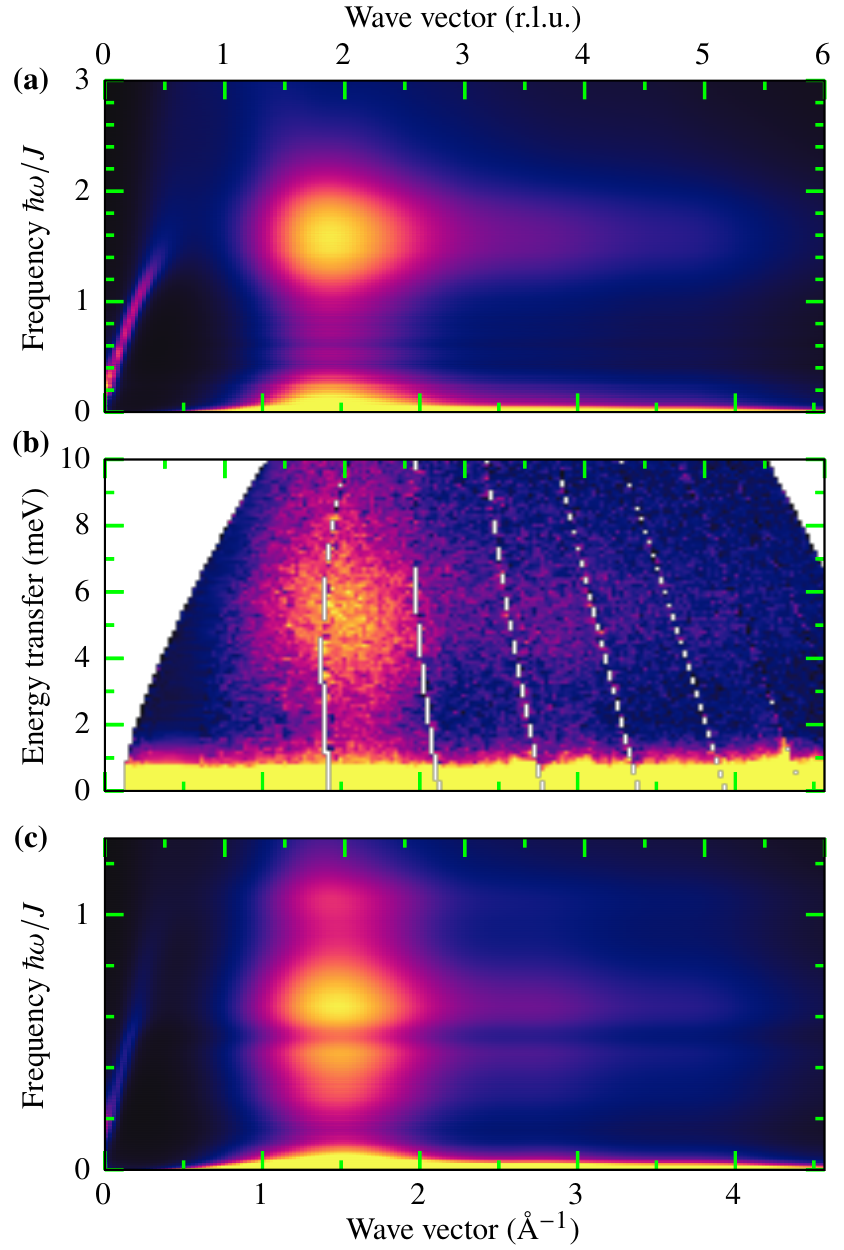}
    \caption{(a) Simulated inelastic neutron-scattering pattern for 10\% Gaussian disorder in $J$ and $Q$ in the BLBQ model [same parameters as Fig.~\ref{fig: broadening}(b)]. 
    (b) Inelastic neutron-scattering intensity of \lgi\ at $T=5.2$~K, measured with 16~meV incident neutron energy~\cite{Tanaka2018InelasticNeutron}.
    (c) Simulated inelastic-neutron scattering pattern for the site-phonon model (same parameters as Fig.~\ref{fig: site phonon}).
    The data in panels (a,c) are multiplied with the Cr$^{3+}$ magnetic form factor, and the wave vector scaled by $2\pi/a_0$ (with lattice parameter $a_0=8.25$~\AA), to aid comparison.
    Frequency ranges are chosen to approximately match the peak positions.
    }
    \label{fig: experiment}
\end{figure}

As shown in Fig.~\ref{fig: experiment}, the theoretically predicted spectrum matches inelastic-neutron-scattering experiments on the nematically ordered spinel \lgi~\cite{Tanaka2018InelasticNeutron}.
Comparing the positions of the inelastic peaks, we estimate $\Qav\approx0.35$~meV assuming the BLBQ model and $\Qav\approx0.75$~meV assuming the site-phonon model.
Taking the length of the $S=3/2$ Cr$^{3+}$ moments into account~\cite{Note11}, the latter corresponds to $\overline{Jb}\approx 0.22$~meV.
Estimates of $J$ and $b$ for \lgi\ in the literature vary widely between $\Jav\approx40$~K~\cite{Okamoto2013BreathingSpinel} to $\Jav\approx80$~K~\cite{gen2023spinlatticecoupled,Ghosh2019BreathingSpinels}.
For the former, our estimate yields $b\approx0.07$, comparable to typical figures in the literature.
On the other hand, the parameters proposed in Ref.~\cite{gen2023spinlatticecoupled} yield $\overline{Jb}\approx 1.0$~meV, almost an order of magnitude too high;
furthermore, their estimate of $J'/J\approx0.04$ leads to low-temperature dynamics desribed by isolated tetrahedra, qualitatively different to the experiment.
Earlier estimates of $J'/J\approx0.6$~\cite{Ghosh2019BreathingSpinels,Okamoto2013BreathingSpinel} appear more consistent with the neutron-scattering results.

The idealised BLBQ spectrum shown in, e.g., Fig.~\ref{fig: powder pattern} differs from the experimental results in  two key ways:
(1) the experimental inelastic peak is far broader than the simulated $16\Qav$ peak;
(2) the linearly dispersing mode is absent in the experiments.
We found that the former can be explained either by finite excitation lifetime due to dynamical processes (modelled using strong Landau--Lifshitz damping $\alpha$) or by static disorder in the Hamiltonian, cf.\ \cref{fig: broadening,fig: experiment}. 
The experimentally observed Gaussian shape of the peak~\cite{Tanaka2018InelasticNeutron}, however, agrees better with the latter scenario.
The additional couplings of the site-phonon model can also be regarded as strong disorder on top of the BLBQ dynamics, which indeed broaden the peak to a similar extent as seen in the experiment.
We believe that the additional structure of this peak would also be washed out by either static disorder or dynamical damping.
In future work, it will be interesting to extend inelastic neutron-scattering measurements to smaller values of $(q,\omega)$, where the fate of the linearly dispersing mode could be studied directly.

Our findings are also potentially relevant to spinel materials beyond \lgi. For example, the solid solution Zn$_{2-x}$Cd$_{x}$Cr$_2$O$_4$ with $x=0.05$ shows a similar phenomenology to \lgi\ in both magnetic susceptibility and inelastic neutron scattering~\cite{Ratcliff2002FreezingZnCdSpinel}, raising the possibility that the nematic state is generic to chromium spinels with light disorder. 
The inelastic excitation energy 4~meV measured for this system~\cite{Ratcliff2002FreezingZnCdSpinel} implies $Q_\mathrm{BLBQ}\approx 0.25$~meV and $Q_\mathrm{sp}\approx 0.55$~meV in the BLBQ and site-phonon models, respectively.
Given the value of $J=3.5$~meV for the end-member ZnCr$_2$O$_4$~\cite{Martinho2001ZnExchange}, the latter implies $b\approx0.04$, consistent with the value $b\approx 0.02$ suggested by high-field magnetisation measurements on ZnCr$_2$O$_4$~\cite{Miyata2012ZnCr2O4}.

We finally note that the order-by-disorder-induced nematic phase of the classical kagome Heisenberg model~\cite{Taillefumier2014SemiclassicalLattice} also exhibits sharp linearly dispersing dynamical modes on top of a partially ordered nematic background.
The similarity of the ice-like ground states, as well as the dynamics, of these two systems raises the tantalising possibility of a deeper analogy between them.
In future work, therefore, it will be interesting to study the long-time relaxation dynamics of the nematic order in our models: 
In the kagome case, this dynamics is governed by qualitatively different processes from the LSWT-like precession dynamics studied in this work.
The relaxation dynamics may also be affected in exotic ways by kinematic constraints, possibly analogous to the fractal dynamics recently uncovered in pyrochlore spin ice~\cite{Hallen2022DynamicalFractal}.

\begin{acknowledgements}
    We used parts of the NetKet library~\cite{Vicentini2022NetKetSystems} to construct the lattices in our code.
    All heat maps use perceptionally uniform colour maps based on Ref.~\cite{Kovesi2015GoodThem}.
    Computing resources were provided by STFC Scientific Computing Department’s SCARF cluster.
    A.~Sz.\ gratefully acknowledges the ISIS Neutron and Muon Source and the Oxford--ShanghaiTech collaboration for support of the Keeley--Rutherford fellowship at Wadham College, Oxford.
    For the purpose of open access, the authors have applied a Creative Commons Attribution (CC-BY) licence to any author accepted manuscript version arising.
\end{acknowledgements}

\appendix

\section{Details of the dynamical simulations}
\label{app: dynamics}

\subsection{Monte Carlo sampling}

We used Metropolis Monte Carlo with single-spin updates to draw  spin configurations from the thermal ensemble of the Hamiltonian~\eqref{eq: Hamiltonian}. In each step, the proposed spin update was constructed as
\begin{equation}
    \vec s_i' = \frac{\vec s_i + \vec r_i}{|\vec s_i+\vec r_i|},
\end{equation}
where each component of $\vec r_i$ is drawn independently from a Gaussian distribution of variance $\sigma^2= {T/(\Jav+6\Qav)}$, chosen to match the thermal fluctuations under the on-site part of the LSWT Hamiltonian~\eqref{eq: Hamiltonian terms}. 
We note that, since there are no interactions between spins on the same sublattice, the proposal--acceptance cycle can be performed in parallel for all spins of the same sublattice, allowing for efficient vectorisation.

This protocol results in a temperature-independent acceptance rate at low temperatures, indicating that the thermal fluctuations around the ordered state are captured well. 
However, the nematically ordered moments become frozen sufficiently below the ordering temperature, resulting in rather noisy static structure factors from a single run.
Therefore, to obtain the static structure factor shown in Fig.~\ref{fig: hhl plots}(a), we initialised the Monte Carlo with 256 independent spin-ice configurations obtained from a variant of the codes in Refs.~\cite{SpinIceCode,Pearce2022MagneticIridates}.

\subsection{Stochastic dynamics}

To solve the Landau--Lifshitz dynamical equation~\eqref{eq: Landau Lifshitz}, we implemented the semi-implicit integrator SIB proposed in Ref.~\cite{Mentink2010SDE}. This algorithm achieves comparable accuracy to fully symplectic solvers (such as the implicit midpoint method) at a fraction of the computational cost by exploiting the sparseness of the dynamical equation.
We performed 65\,536 time steps of size $\Delta t = \hbar/(16J)$ for a total simulation time $T=4096\hbar/J$, resulting in a frequency resolution $\Delta\omega = 2\pi/T \approx1.53\times10^{-3} J/\hbar$. We only saved the spin configuration after every fourth step, as this still allowed us to resolve the full dynamical spectrum.

In stochastic differential equation solvers, the noise term $\vec b_i$ of~\eqref{eq: Landau Lifshitz} is implemented using a noise vector $\vec\xi_i$, whose components should be unit Gaussian random numbers.
Due to the implicitness of the solver, however, using unbounded $\vec\xi_i$ can lead to instabilities.
Ref.~\cite{Mentink2010SDE} proposes to simply apply a cutoff to Gaussian $\vec\xi_i$ components: we found that this results in strong numerical damping and equilibrium energies well below that obtained for the same temperature from Monte Carlo at any temperature, time step size, or value of $\alpha$.
By contrast, drawing $\vec\xi_i$ uniformly from the surface of a sphere of radius $\sqrt3$ (such that the standard deviation of each component is 1) resulted in energies that match the Monte Carlo results within statistical error. We believe that matching the (co)variances of the ideal Gaussian noise in any projection (not only along the Cartesian axes) is crucial for this.

For most parameter values, we ran a single dynamical simulation, as the dynamical fluctuations appear to remain self-averaging even in a frozen spin-ice background. For the parameters $J=J',Q=Q'=0.1J,\alpha=0.01$ used in \cref{fig: powder pattern,fig: hhl plots,fig: lswt intensity}, we averaged four independent runs to improve statistics.

\subsection{Powder averaging}
To compute powder averages of $q$-dependent quantities, we broadened every $k$-point obtained from FFT with a Gaussian of standard deviation
$\sigma_q = {\sqrt{2\pi}}/{La_0}$,
where $L$ is the number of cubic unit cells along each Cartesian direction ($La_0$ is the linear size of the simulation box) and integrated the result over bins of width $\Delta q$:
\begin{align}
    S_\mathrm{powder}(q) &= \frac 1{4\pi q^2\Delta q}\int_{q-\Delta q/2}^{q+\Delta q/2} \ud q' \sum_k S(\vec k) \frac{e^{-(k-q')^2/2\sigma_q^2}}{\sqrt{2\pi} \sigma_q}\nonumber\\
    &= \frac 1{4\pi q^2\Delta q}\sum_k \frac{S(\vec k)}2 \left[\erf\left(\frac{q' -k}{\sqrt 2\sigma_q}\right)\right]_{q-\Delta q/2}^{q'=q+\Delta q/2}.
\end{align}
We can think of this as spreading out the discrete $k$-points into three-dimensional Gaussians in reciprocal space and averaging the result over spherical shells; the denominator $4\pi q^2\Delta q$ is the volume of such a shell.
The width $\sigma_q$ of the Gaussian was chosen such that the effective volume $(\sqrt{2\pi}\sigma_q)^3$ taken up by them in reciprocal space match the volume around each allowed $k$-point, $(2\pi/La_0)^3$. Both this choice and integrating over bins of $q$ reduce spurious fluctuations due to the discrete $k$-points available on the finite-size system, allowing us to use the relatively narrow bin width $\Delta q=\pi/La_0$.

\subsection{Dynamical structure factor of the site-phonon model}
\label{app: site phonon gallery}

\begin{figure*}
    \centering
    \includegraphics{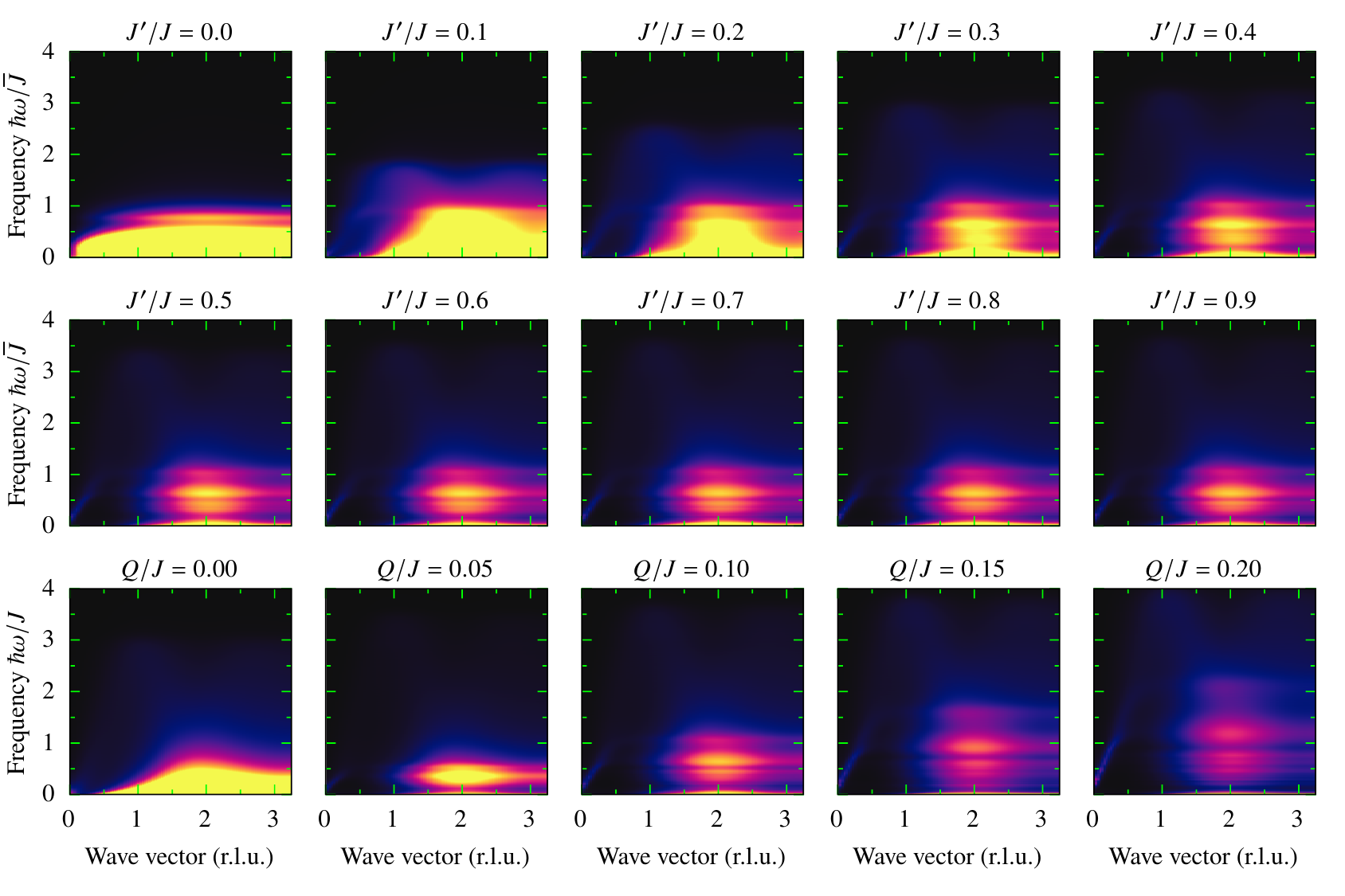}
    \caption{Simulated powder-averaged dynamic structure factor of the site-phonon model~\eqref{eq: site-phonon Hamiltonian} for the same parameters as used in Fig.~\ref{fig: all powder patterns}.}
    \label{fig: site phonon tour}
\end{figure*}

We performed dynamical simulations of the site-phonon model~\eqref{eq: site-phonon Hamiltonian} for all the parameter sets used in Fig.~\ref{fig: all powder patterns}.
The resulting dynamical structure factors are shown in Fig.~\ref{fig: site phonon tour}.
Similar to the BLBQ model, we find that the finite-frequency peak and the linearly dispersing mode form clearly for all $J'/J\gtrsim 0.3$, and all quantitative features are essentially the same for all $J'\gtrsim 0.5$.
The position of the main inelastic peak also appears to remain proportional to $Q$.

\section{Derivation of the LSWT dynamical equation}
\label{app: LSWT derivation}

Eq.~\eqref{eq: Landau Lifshitz linearised} can be obtained by straightforwardly expanding~\eqref{eq: Landau Lifshitz} to first order in $s^\pm$. Here, we present an alternative derivation that makes explicit use of the quadratic Hamiltonian~\eqref{eq: Hamiltonian quadratic} and thus explains the presence of the matrix $H$ in~\eqref{eq: Landau Lifshitz linearised}.

The energy-conserving dynamical term $\vec s_i\times \vec B_i$ can be obtained by applying Ehrenfest's theorem to~\eqref{eq: Hamiltonian} and replacing every spin operator with its expectation value~\cite{Szabo2019SeeingIce}. Likewise, the semiclassical dynamical equation for $s_i^+$ is
\begin{align}
    \frac{\ud s_i^+}{\ud t} &= \frac{\ud \langle \hat s^+_i\rangle}{\ud t} 
    = -\frac{i}{\hbar} \langle [\hat s^+_i, \mathcal H]\rangle 
    \approx -\frac{i}{\hbar} \left\langle \left[\hat s^+_i, \frac12 \hat s_j^+ H_{jk} \hat s_k^-\right]\right\rangle \nonumber\\
    &= -i \langle \hat s_j^+ H_{ji} \hat s_i^z\rangle \approx -\frac{i}\hbar  S_i H_{ij} s_j^+,
    \label{eq: conservative dynamics Ehrenfest}
\end{align}
where we also used that $\langle \hat s_i^z\rangle = S_i$ to leading order.

There is no equally straightforward derivation of the dissipative term from first principles. However, the $z$-component of $\vec s_i\times \vec B_i$ comes from the transverse components of $\vec s_i,\vec B_i$, so it is of second order. Therefore, the only first-order terms in $\vec s_i \times \vec s_i\times \vec B_i$ are due to the $z$-component of $\vec s_i$ and the transverse components of $\vec s_i \times \vec B_i$:
\begin{align}
    (\vec s_i \times \vec s_i\times \vec B_i)^+ &= s_i^z \big[-(\vec s_i\times \vec B_i)^y + i(\vec s_i\times \vec B_i)^x\big] \nonumber\\
    &= iS_i (\vec s_i\times \vec B_i)^+ = \frac1\hbar S_i^2 H_{ij} s_j^+,
    \label{eq: dissipative term}
\end{align}
where we substitute~\eqref{eq: conservative dynamics Ehrenfest} for $(\vec s_i\times \vec B_i)^+$. Substituting (\ref{eq: conservative dynamics Ehrenfest},\,\ref{eq: dissipative term}) into~\eqref{eq: Landau Lifshitz} and using that $S_i^2 = 1$ yields~\eqref{eq: Landau Lifshitz linearised}.

\section{Mathematical properties of the LSWT equations}
\label{app: LSWT matrices}

We expect that $H$ be a positive (semi)definite matrix when performing the expansion~\eqref{eq: Hamiltonian quadratic} near an Ising configuration that obeys the ice rules, as these minimise the Hamiltonian~\eqref{eq: Hamiltonian}.
We can show this mathematically by rewriting $H$ as a sum of terms of the form~\eqref{eq: Hamiltonian on one tetrahedron} acting on individual tetrahedra. The eigenvalues of these terms are 0 (for the mode locally proportional to $S_i$), $4J+8Q$ [for the ice-rule-violating mode $(1,1,1,1)$], and $8Q$ (for the two modes orthogonal to both of these). 
That is, each term is positive semidefinite, so $\langle v|H|v\rangle\ge 0$ for all $|v\rangle$ as well, and zero modes must be zero modes of every term. 
For $Q > 0$ therefore, the only zero mode of $H$ is that proportional to $S_i$, i.\,e., rigid rotations of the spin-ice configuration.
For $Q = 0$, every mode that respects the ice rule has zero energy, which explains the breakdown of the linear-spin-wave picture for the pure Heisenberg model.

For $\alpha=0$, the Landau--Lifshitz dynamics~\eqref{eq: Landau Lifshitz} conserves energy. Near an energy minimum, this prevents the linear spin-wave dynamics from having any exponentially decaying or exploding modes, thus all eigenvalues of the dynamical matrix $SH$ must be real. To prove that this is the case, we multiply~\eqref{eq: lswt eigenvalue eq} with $H$ from the left:
\begin{equation}
    HSH|r_a\rangle = \hbar\omega_a H|r_a\rangle.
\end{equation}
Now, both $HSH$ and $H$ are Hermitian matrices, and $H$ is positive definite~%
\footnote[101]{The zero mode of $H$ could cause problems. However, it is obviously a zero mode of $SH$ too, so we could remove it from the basis and work with a positive definite $H$. 
Note also that this (semi)definiteness relies on the reference configuration being a minimum of $\mathcal H$: around a general Ising configuration, the dynamical matrix will have complex eigenvalues corresponding to instabilities near ice-rule-violating tetrahedra.}:
this implies that the eigenvalues $\omega_a$ are all real~\cite{Parlett1998Eigenvalues}. 
As both matrices are also real, the eigenvectors are real, too.

We can extend these arguments to show that the eigenfrequencies of the $\alpha>0$ dynamics have $\Im\omega\le 0$, that is, they all decay. We multiply~\eqref{eq: lswt eigenvalue eq} from the left by $H^{1/2}$ to get
\begin{align}
    H^{1/2}(S-i\alpha)H^{1/2} |\tilde r_a\rangle &= \hbar\omega_a|\tilde r_a\rangle. &
    (|\tilde r_a\rangle &\equiv H^{1/2} |r_a\rangle)
    \label{eq: symmetrised eigenvalue eq}
\end{align}
Multiplying on the left by $\langle\tilde r_a|$, we find that the eigenfrequency is given by the Rayleigh quotient
\begin{align}
    \hbar\omega &= \frac{\langle\tilde r_a| H^{1/2}(S-i\alpha) H^{1/2}|\tilde r_a\rangle}{\langle\tilde  r_a|\tilde r_a\rangle}\nonumber\\
    &= \frac{\langle r_a|HSH|r_a\rangle}{\langle r_a|H|r_a\rangle} - i\alpha \frac{\langle r_a|H^2|r_a\rangle}{\langle r_a|H|r_a\rangle}.
    \label{eq: Rayleigh quotient}
\end{align}
Every matrix in the second form of~\eqref{eq: Rayleigh quotient} is Hermitian, so $\Im\omega$ is entirely due to the second term. Furthermore, both $H$ and $H^2$ are positive definite~\cite{Note101}, so ${\langle r_a|H^2|r_a\rangle}/{\langle r_a|H|r_a\rangle} > 0$ for any $|r_a\rangle$, thus $\Im\omega_a <0$ for all but the zero mode.

The matrix $H^{1/2}(S-i\alpha)H^{1/2}$ in~\eqref{eq: symmetrised eigenvalue eq} is symmetric. Complex symmetric matrices are generally not normal, so their left and right eigenvectors are different. However, taking the transpose (\textit{not} conjugate transpose) of~\eqref{eq: symmetrised eigenvalue eq} gives
\begin{equation}
    \langle \tilde r_a^*|H^{1/2}(S-i\alpha)H^{1/2} = \langle \tilde r_a^*| \hbar\omega,
\end{equation}
that is, $\langle \tilde r_a^*|$ is a left eigenvector corresponding to the same eigenvalue as $|\tilde r_a\rangle$. The resolvent of $H^{1/2}(S-i\alpha)H^{1/2}$ can therefore be written as
\begin{equation}
    [\hbar\omega - H^{1/2}(S-i\alpha)H^{1/2}]^{-1} = \sum_a \frac{|\tilde r_a\rangle \langle \tilde r_a^*|}{\hbar(\omega-\omega_a)},
    \label{eq: resolvent 1}
\end{equation}
assuming the usual orthonormalisation for left and right eigenvectors,
\begin{equation}
    \langle \tilde r_a^*|\tilde r_b\rangle = \langle r_a^*| H|r_b\rangle = \delta_{ab}.
    \label{eq: orthogonality}
\end{equation}
Finally, multiplying~\eqref{eq: resolvent 1} on both sides by $H^{-1/2}$ gives
\begin{equation}
    [\hbar\omega H - H(S-i\alpha)H]^{-1} = \sum_a \frac{|r_a\rangle \langle r_a^*|}{\hbar(\omega-\omega_a)}.
    \label{eq: resolvent}
\end{equation}

\section{Dynamical structure factor from LSWT}
\label{app: LSWT dynamical structure factor}

To incorporate the stochastic fluctuation terms of~\eqref{eq: Landau Lifshitz} into linear spin-wave theory, we note that the leading-order components of $\vec s_i\times\vec b_i$ and $\vec s_i\times\vec s_i\times\vec b_i$ are those involving the $z$-component of $\vec s_i$. Therefore,
\begin{align}
    \hbar \frac{\ud |s^+\rangle}{\ud t} &= -(iS+\alpha)H|s^+\rangle + (iS+\alpha)|b^+\rangle \nonumber\\
    -i\hbar \omega |s^+(\omega)\rangle &= -(iS+\alpha)H|s^+(\omega)\rangle +(iS+\alpha)|b^+(\omega)\rangle\nonumber\\
    |s^+(\omega)\rangle &= [-i\hbar\omega+(iS+\alpha)H]^{-1}(iS+\alpha)|b^+(\omega)\rangle \nonumber\\
    &= -[\hbar\omega-(S-i\alpha)H]^{-1} (S-i\alpha)|b^+(\omega)\rangle,
\end{align}
where $|b^+\rangle$ is the vector of $b_i^x+ib_i^y$.

The dynamical structure factor in real space is given by the thermal average of $|s^+(\omega)\rangle\langle s^+(\omega)|$. To perform the average, we note that
\begin{equation}
    \langle b_i^+(t) b_j^-(t')\rangle = 4D \delta_{ij} \delta(t-t') \implies
    \langle b_i^+(\omega) b_j^-(-\omega)\rangle = 4D \delta_{ij},
\end{equation}
whence
\begin{widetext}
\begin{align}
    \mathcal{S}(\omega) = \Big\langle |s^+(\omega)\rangle \langle s^+(\omega)|\Big\rangle 
    &= [\hbar\omega - (S-i\alpha)H]^{-1} (S-i\alpha) 4D (S+i\alpha) [\hbar\omega - H(S+i\alpha)]^{-1} \nonumber\\
    &= 4D(1+\alpha^2) [\hbar\omega - (S-i\alpha)H]^{-1} [\hbar\omega - H(S+i\alpha)]^{-1} \nonumber\\
    &= 4D(1+\alpha^2) [\hbar\omega - (S-i\alpha)H]^{-1} H  [\hbar\omega - (S+i\alpha)H]^{-1} H^{-1} \nonumber\\
    &= \frac{2iD(1+\alpha^2)}{\alpha} \left\{[\hbar\omega - (S-i\alpha)H]^{-1} - [\hbar\omega - (S+i\alpha)H]^{-1}\right\} H^{-1} \nonumber\\
    &= 2ik_\mathrm{B}T\hbar \left\{[\hbar\omega H - H(S-i\alpha)H]^{-1} - [\hbar\omega H - H(S+i\alpha)H]^{-1}\right\} \nonumber\\
    &= 2ik_\mathrm{B} T \sum_a \left(\frac1{\omega-\omega_a} \frac{| r_a\rangle \langle r_a^*|}{\langle r_a^*|H| r_a\rangle} - \frac1{\omega-\omega_a^*} \frac{| r_a^*\rangle \langle r_a|}{\langle r_a|H| r_a^*\rangle} \right).
    \label{eq: dynamical structure factor matrix}
\end{align}
\end{widetext}
In the second line, we use that $S$ is a diagonal matrix with $\pm1$ as entries.
The fourth line uses the identity $(A+B)^{-1} - (A-B)^{-1} = -2 (A+B)^{-1} B (A-B)^{-1}$. In the last two lines, we substitute the fluctuation--dissipation relation~\eqref{eq: fluctuation dissipation} and the spectral decomposition~\eqref{eq: resolvent}, making the normalisation~\eqref{eq: orthogonality} explicit. The two terms in the last two lines are manifestly complex conjugate symmetric matrices, so $\mathcal S(\omega)$ is real and symmetric, as expected from its definition.

Eq.~\eqref{eq: dynamical structure factor matrix} assumes that the eigenvectors $|r_a\rangle$ satisfy the orthogonality condition~\eqref{eq: orthogonality}. For degenerate modes (such as the exact $16\Qav$ modes), the eigenvectors returned by nonhermitian eigensolvers do not satisfy any such relation, so blindly applying~\eqref{eq: dynamical structure factor matrix} leads to incorrect results. For the results shown in Fig.~\ref{fig: lswt intensity}, we added very weak ($\Delta J/J =10^{-6}$) bond disorder to lift all degeneracies without perceptibly changing $\mathcal{S}(\omega)$.

The $q$-integrated and $q$-resolved dynamical structure factors can be expressed from~\eqref{eq: dynamical structure factor matrix} as
\begin{align}
    \frac1N \tr \mathcal{S}(\omega) &= -\frac{4k_\mathrm{B} T}N \sum_a \Im\left(\frac1{\omega-\omega_a} \frac{ \langle r_a^*|r_a\rangle}{\langle r_a^*|H| r_a\rangle}\right) ;\\
    \langle q|\mathcal{S}(\omega) |q\rangle &= -4k_\mathrm{B} T  \sum_a \Im\left(\frac1{\omega-\omega_a} \frac{\langle q| r_a\rangle \langle r_a^*|q\rangle}{\langle r_a^*|H| r_a\rangle}\right) \nonumber\\*
    &= -4k_\mathrm{B} T  \sum_a \Im\left(\frac1{\omega-\omega_a} \frac{r_a(q) r_a(-q)}{\langle r_a^*|H| r_a\rangle}\right),
\end{align}
where $r_a(q)$ are the Fourier components of the eigenvector.
Static structure factors can be obtained from these by integrating over $\omega$ and noting that 
\begin{equation*}
    \mathcal{P} \int_{-\infty}^\infty \frac{\ud\omega}{2\pi} \frac1{\omega-\omega_a} = \frac{i}2\sgn(\Im\omega_a) = -i/2
\end{equation*}
for every $\omega_a$ in the lower half plane. In particular, the mean square transverse fluctuation of each spin is given by
\begin{align}
    \langle s_i^+ s_i^-\rangle &= \frac1N \int_{-\infty}^\infty \frac{\ud \omega}{2\pi} \tr \mathcal{S}(\omega) = \frac{2k_\mathrm{B}T}N \sum_{a\neq 0} \Re \frac{ \langle r_a^*|r_a\rangle}{\langle r_a^*|H| r_a\rangle};
    \label{eq: static trace}
\end{align}
the zero mode must be excluded as it does not correspond to transverse fluctuations around a nematic state but rotating its director.

We finally note that in the limit $\alpha\to 0^+$, $1/(\omega-\omega_a) \to \mathcal{P}1/(\omega-\omega_{0a}) - i\pi \delta(\omega-\omega_{0a})$, so~\eqref{eq: dynamical structure factor matrix} becomes
\begin{equation}
    \mathcal{S}(\omega) = 2k_\mathrm{B}T \sum_a \frac{|r_a\rangle\langle r_a|}{\langle r_a|H|r_a\rangle} 2\pi\delta(\omega-\omega_{0a}).
\end{equation}
That is, each eigenmode of $SH$ gives rise to a sharp peak in the structure factor, with a spatial structure matching the eigenvector $|r\rangle$ and intensity normalised by the energy cost $\langle r|H|r\rangle$ of exciting the mode.

\section{Quadratic expansion of the site-phonon Hamiltonian}
\label{app: site phonon quadratic}

\begin{figure}
    \centering
    \includegraphics{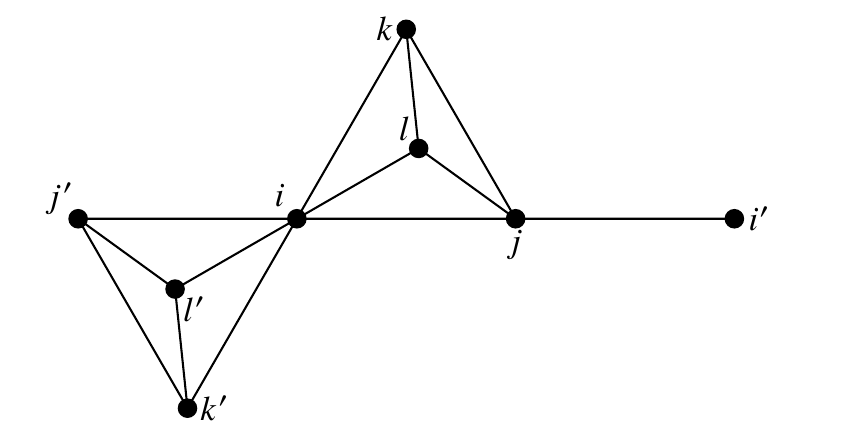}
    \caption{Layout of site labels used in~\eqref{eq: site phonon quadratic generic}.}
    \label{fig: site labels}
\end{figure}

Substituting~\eqref{eq: LSWT expansion} into $\vec s_i\cdot \vec s_j$ and expanding to second order in $s^\pm$ yields
\begin{equation}
    \vec s_i\cdot\vec s_j \simeq S_iS_j + \frac12 \underbrace{\left[ s_i^+ s_j^- + s_j^+ s_i^- - S_iS_j ( s_i^+s_i^- + s_j^+s_j^-)\right]}_{b_{ij}}.
\end{equation}
Since this has no linear term in $s^\pm$, all quadratic terms in the expansion of any $(\vec s_i\cdot\vec s_j)(\vec s_k \cdot\vec s_l)$ contain the zeroth-order term of one $\vec s\cdot\vec s$ and the quadratic term of the other. 
In particular, the quartic term of the site-phonon Hamiltonian~\eqref{eq: site-phonon Hamiltonian} becomes
\begin{align}
    H_{\mathrm{sp},Q} &\simeq -\frac12\sum_i \sum_{j,k\sim i} \sqrt{Q_{ij}Q_{ik}} (\hat e_{ij} \cdot \hat e_{ik}) \frac{S_iS_j b_{ik} + S_iS_k b_{ij}}2 \nonumber\\
    &= -\frac12 \sum_i \sum_{j,k\sim i} \sqrt{Q_{ij}Q_{ik}} (\hat e_{ij} \cdot \hat e_{ik}) S_iS_k b_{ij}
    \label{eq: site phonon quadratic 1}
\end{align}
up to an overall constant.
Since the only quadratic terms come from expanding nearest-neighbour $\vec s_i\cdot\vec s_j$, Eq.~\eqref{eq: site phonon quadratic 1} is still nearest-neighbour.
The expanded BLBQ Hamiltonian~\eqref{eq: Hamiltonian terms} can be recovered from~\eqref{eq: site phonon quadratic 1} by keeping the $j=k$ terms only.

Let us now consider the coefficients of $b_{ij}$ in~\eqref{eq: site phonon quadratic 1} where $i$ appears in the outer sum.
For convenience, we assume that the bond $ij$ is on an up tetrahedron; for a down tetrahedron, $Q\leftrightarrow Q'$.
Writing out the sum over $k$ explicitly, we get
\begin{align}
    H_{\mathrm{sp},Q}&\supset \frac{b_{ij}}2\left[-Q \left(S_iS_j +\frac12 S_iS_k+\frac12 S_iS_l\right)\right. \nonumber\\*
    & \hspace*{4em}\left. {}+ {\sqrt{QQ'}}  \left(S_iS_{j'} +\frac12 S_iS_{k'}+\frac12 S_iS_{l'}\right)\right]
    \label{eq: site phonon quadratic generic} \\
    &= \frac{b_{ij}}2\left[ -\frac{Q}2 S_iS_j + \frac{Q}2 + \frac{\sqrt{QQ'}}2 S_iS_{j'} - \frac{\sqrt{QQ'}}2\right],
    \label{eq: site phonon quadratic 2}
\end{align}
where the site labels are as shown in Fig.~\ref{fig: site labels}. 
In~\eqref{eq: site phonon quadratic 2}, we also assume that the $S_i$ form a spin-ice configuration such that $S_j+S_k+S_l = S_{j'}+S_{k'}+S_{l'} = -S_i$.
After including the terms where $j$ appears in the outer sum, we get the full coefficient of $b_{ij}$:
\begin{equation}
    H_{\mathrm{sp},Q}\supset \frac{b_{ij}}2 \left[ ({Q-\sqrt{QQ'}}) - QS_iS_j + \sqrt{QQ'} \frac{S_iS_{j'}+S_jS_{i'}}2 \right].
\end{equation}

\bibliography{references,paper}

\end{document}